\documentclass[conference]{IEEEtran}

\AtBeginDocument{%
  }

\usepackage{amsmath,amssymb,amsfonts}
\usepackage{algorithmic}
\usepackage{graphicx}
\usepackage{textcomp}
\usepackage{xcolor}
\usepackage{svg}
\usepackage{booktabs} 
\usepackage{adjustbox}
\usepackage{url}
\begin{document}

\title{Compiler and Hardware Co-Design for Accelerator Architectures\\
}

\author{\IEEEauthorblockN{Karl Herman Krause}
\IEEEauthorblockA{\textit{Department of Applied Mathematics and Computer Science} \\
\textit{Technical University of Denmark}\\
Lyngby, Denmark \\
s203852@student.dtu.dk}
\and
\IEEEauthorblockN{Emad Jacob Maroun}
\IEEEauthorblockA{\textit{Department of Applied Mathematics and Computer Science} \\
\textit{Technical University of Denmark}\\
Lyngby, Denmark \\
ejama@dtu.dk}
\and
\IEEEauthorblockN{Martin Schoeberl}
\IEEEauthorblockA{\textit{Department of Applied Mathematics and Computer Science} \\
\textit{Technical University of Denmark}\\
Lyngby, Denmark \\
masca@dtu.dk}
}

\newcommand{\martin}[1]{{\color{blue} Martin: #1}}
\newcommand{\emad}[1]{{\color{red} Emad: #1}}
\newcommand{\karl}[1]{{\color{violet} Karl: #1}}


\maketitle


\begin{abstract}
Heterogeneous accelerator architectures offer an efficient path to performance for compute-intensive workloads. 
However, full-stack integration remains difficult. We present EAAC (Extensible Accelerator Architecture), a flexible and extensible compiler and hardware architecture designed to lower the overhead of hardware-compiler co-development for rapid prototyping of hardware accelerators. EAAC targets static data-flow workloads with predictable memory access patterns.

By using MLIR, we enable possible integration with a range of different frontends that emit MLIR. And by providing a minimal set of compiler functionality that enable necessary data-orchestration we lower the effort needed to get a simple implementation of a hardware acceleration unit up and running, while providing a relatively blank and un-opinionated starting point for further work.

We validate this approach with a prototype GEMM accelerator combining a systolic array unit with an embedded RISC-V core, compiled end-to-end through the EAAC MLIR pipeline. On a synthetic fully-connected-layer workload, the accelerator achieves a 28x speedup in execution time over a RISC-V-only baseline, with the GEMM operation itself accounting for only 1.64\% of total execution time. We further characterize the compiler's hardware-semaphore allocation, showing that the number of semaphores required scales linearly with instruction count in the worst case, and identify this as a concrete target for future optimization.
\end{abstract}

\section{Introduction}

As modern workloads evolve, domain-specific hardware acceleration has become an increasingly important solution for handling computationally demanding tasks. With memory access consuming disproportionately more time and energy than computation in modern architectures, hardware accelerators excel in applications with predictable memory access patterns that enable greater data reuse \cite{10.1145/3361682}. While predictable memory access has long been associated with good performance on traditional multicore CPUs, the combination of massive parallelism and improved memory efficiency in accelerator architectures results in substantially higher performance and energy efficiency.
A consequence of the high degree of specialization of these devices is that the surrounding software infrastructure cannot be assumed to exist and must be developed alongside the hardware itself. This presents a significant barrier to entry for teams with limited resources, delays full-stack testing, and slows iteration even for larger teams \cite{article}.
Existing frameworks address this primarily by raising the abstraction level of accelerator implementation through high-level synthesis (HLS), accelerator design languages, or hardware compiler infrastructures such as Allo \cite{10.1145/3656401}, Calyx \cite{DBLP:journals/corr/abs-2102-09713}, and CIRCT \cite{circt-github}. While these tools lower the barrier to hardware implementation, they do not directly address the cost of integrating the resulting accelerator into a full software stack, and typically produce a fixed hardware instantiation tied to a specific program or workload.
We propose EAAC (Extensible Accelerator ArChitecture), a framework for hardware-compiler co-development focused on reducing the cost of hardware/software integration. Rather than raising the abstraction level of hardware implementation, EAAC defines a minimal and flexible interface between accelerator hardware units and the surrounding system. EAAC is specifically designed for workloads that are optimal for acceleration, often characterized by regular and predictable memory access patterns, minimal control flow, and statically analyzable data dependencies.
EAAC is not intended to automate the hardware or compiler development process, but rather to provide a flexible abstraction that can function as a starting point for further development. The EAAC compiler is capable of static memory allocation and manages a multi-tiered memory hierarchy using explicit direct memory access (DMA) control between layers, and additionally introduces data-driven synchronization 
using hardware semaphores. This allows accelerators produced through different methods, including hand-written RTL, Chisel generators, \cite{SCHOEBERL2025105182}
and HLS-generated designs, to coexist within a shared framework, while unifying data, control, and status 
interfaces between hardware units into a single bus, by utilizing memory-mapped hardware semaphores for control and synchronization.
This paper makes the following contributions:
\begin{itemize}
  \item A flexible hardware architecture based on decoupled access-execute \cite{800048.801719}, with phased execution allowing for data transfers to overlap with execution.
  \item An MLIR (multi-Level intermediate representation) compiler pipeline with static memory allocation and explicit synchronization for data-driven scheduling.
  \item A prototype general matrix multiplication (GEMM) accelerator implemented using the EAAC framework, serving as a proof-of-concept evaluation target.
\end{itemize}

The code for the EAAC compiler, ATAN hardware accelerator (Prototype accelerator) and MCC (minimal control core, RISC-V) are available on github.

\begin{itemize}
    \item https://github.com/McHerman/eaac-dialect
    \item https://github.com/McHerman/ATAN
    \item https://github.com/McHerman/mcc
\end{itemize}

\section{Related work}

Act \cite{jain2025actautomaticallygeneratingcompiler} presents the same fundamental issues as we address in this paper, but proposes a different solution by generating the compiler backend for a tensor processing hardware accelerator from a formal description of its instruction set architecture (ISA). Act dissolves tensor operations into a set of IR-like primitive tensor operations; ISA instructions are then defined in terms of these primitives, and hardware kernels can, in turn, be defined in terms of those instructions. The system handles memory allocation and data movement through integer constraint programming.

EAAC, in comparison, has a much more conventional sequential pass-based compiler. EAAC and Act both treat tensors as first-class datatypes, but while Act decomposes all arithmetic operations into a combination of tensor-modification primitives, EAAC's tensor semantics stem mainly from applying bufferization to tensors during memory allocation; as a result, any operation using statically allocated memrefs can be compiled for EAAC.

SNAX \cite{antonio2025opensourcehwswcodevelopmentframework} is another project that EAAC mirrors closely, both in intention and implementation, but with some significant and relevant differences. SNAX utilizes RISC-V processor cores for synchronization, configuration, and control of the hardware accelerator units, specifically through control and status registers in the RISC-V core, which map to a standardized interface for the accelerator units. The system therefore creates a direct connection between the RISC-V core and the accelerator units, via an interface that is ``tightly'' integrated into the fabric of the accelerator.

EAAC utilizes a distributed control system that does not place a single unit in the role of an orchestrator; rather, hardware units communicate directly using hardware semaphores. While EAAC does support offloading to RISC-V processors embedded into the accelerator architecture, it does not depend on the RISC-V for scheduling and issuing commands to other accelerator units in the system. EAAC and SNAX use comparable MLIR-based compilers; both snax-mlir and EAAC utilize MiniMalloc \cite{10.1145/3623278.3624752} for static memory allocation.

Many HLS projects attempt to solve the difficulties of implementing hardware accelerators by generating the compiler and hardware architecture from the same source. Tools such as Allo \cite{10.1145/3656401}, Spatial \cite{10.1145/3192366.3192379}, and Calyx \cite{DBLP:journals/corr/abs-2102-09713} function as programming models that separate the scheduling and hardware configuration from the definition of a given algorithm or software kernel, but the hardware they generate is hard-coded for a given application, making them well suited for quick iteration and deployment on FPGAs but unsuited for ASICs. Coarse-Grained Reconfigurable Arrays (CGRAs) \cite{DBLP:journals/corr/abs-2004-04509} occupy a different point in the design space: like FPGAs, they expose a reconfigurable interconnect, but their functional units resemble small processing elements rather than configurable logic blocks, making CGRAs more area- and power-efficient than FPGAs at the cost of some flexibility. AHA \cite{10.1145/3534933} is a hardware/software co-design tool that targets CGRAs specifically, generating both hardware and software from the same source while allowing fine-tuning of the individual processing elements to the degree of application-specific reconfiguration required; it further compiles applications written in the Halide programming language to the generated hardware, giving it a degree of re-programmability for a fixed hardware design that the HLS tools above do not offer. Broadly, HLS and CGRAs sit at different ends of a flexibility-versus-efficiency spectrum: HLS flows can approach the efficiency of hand-optimized accelerators for fixed workloads, while CGRAs trade some peak efficiency for the ability to support a wider range of programs \cite{DBLP:journals/corr/abs-2004-04509}. EAAC is designed to occupy a position between these extremes: more flexible than a fixed HLS-generated accelerator, but more domain-focused and potentially more efficient than a general-purpose CGRA, at the cost of requiring more integration effort when adapting to a new workload.

\section{Background}

\subsection{MLIR}

Domain-specific hardware acceleration has historically complicated compiler backend development, as the compiler must be built for both a specific workload and a specific hardware architecture, often requiring significant development effort from the ground up \cite{DBLP:journals/corr/abs-2002-11054}. This is a significant burden on the development of new and esoteric architectures. The popularity of architectures like RISC-V is partly driven by the fact that a shared ISA allows many different micro-architectures to share a common compiler toolchain \cite{10049118}. Similar development within the hardware acceleration space has been slower, due to the natural fragmentation of the field. 

MLIR \cite{DBLP:journals/corr/abs-2002-11054} is a compiler framework designed with heterogeneous and domain-specific computing in mind. Its intermediate representation is more flexible than LLVM IR \cite{llvm-langref} and supports user-extensible dialects.
Developers can define custom IR operations and transformation passes to support compiling for targets other than classical assembly languages. 
This extensibility makes MLIR a suitable foundation for compiler infrastructure targeting novel hardware architectures \cite{10.1145/3706628.3708870}, \cite{11409379}.

MLIR dialects represent different abstractions: they can represent the ISA instructions belonging to a specific hardware architecture, such as the \textit{amdgpu} dialect \cite{mlir-amdgpu-dialect}, which functions as a wrapper around AMD GPU instructions.
They can also represent domain-specific operations that do not belong to any specific hardware architecture, such as the \textit{linalg} dialect used to represent different tensor operations.

MLIR is intended to function as an intermediate representation supported by multiple frontends. One popular application of MLIR is as a compiler for ML workloads. Different ML frameworks such as TensorFlow and PyTorch allow for exporting models into MLIR using different dialects that represent typical ML primitives such as general matrix multiplication (GEMM) and various tensor types.

To convert tensors into a memory representation that is closer to hardware, bufferization on tensors is used. It converts tensor types to MLIR MemRefs. Memrefs are MLIR's core dialect for memory references, and act as pointers to buffers in memory with additional information about element types, shapes, and memory space \cite{mlir-memref-dialect}. Using bufferization on tensors, MLIR generates IR (Intermediate Representation) explicitly declaring memory allocations and deallocations.

%
%
\subsection{Target Workload Characteristics}

Different hardware architectures naturally call for different compiler strategies. EAAC is specifically designed with static data-flow applications in mind. These properties are relevant because they enable memory access scheduling optimization in the compiler, including data pre-fetching through buffer lifetime analysis and dependency tracking, which in turn allows for a simpler and more efficient hardware design \cite{1458143}. While many workloads can benefit from some degree of hardware acceleration, this class of workloads is particularly well-suited to a statically scheduled, compiler-managed execution model \cite{10.1145/3361682}.

Many of the applications that have typically warranted the investment in a domain-specific hardware accelerator have these traits \cite{PECCERILLO2022102561}. Examples include GEMM, digital signal processing, and image processing \cite{PECCERILLO2022102561}.


\section{Extensible Accelerator Architecture}

EAAC is a hardware-software co-development framework that aims to provide a flexible and practical starting point for custom hardware accelerator development. This is achieved primarily by exposing a unified data-orchestration framework that wraps around custom hardware acceleration units, while providing software control over memory and dependency management.

EAAC operates exclusively using data-driven scheduling; unlike a typical von Neumann architecture, it does not have a program counter, branching, or looping. The execution of the machine is represented entirely as a stream of instructions issued to it sequentially. This stream can originate from an embedded RISC-V core or from a host device through a bus interface, as in the case of the prototype architecture explained later in this paper.

\subsection{Hardware Architecture}

EAAC is a system built to support a diverse range of hardware units for different acceleration purposes, but it fundamentally adheres to a single abstract architecture.

\begin{figure}
\centering
\includegraphics[scale=0.50]{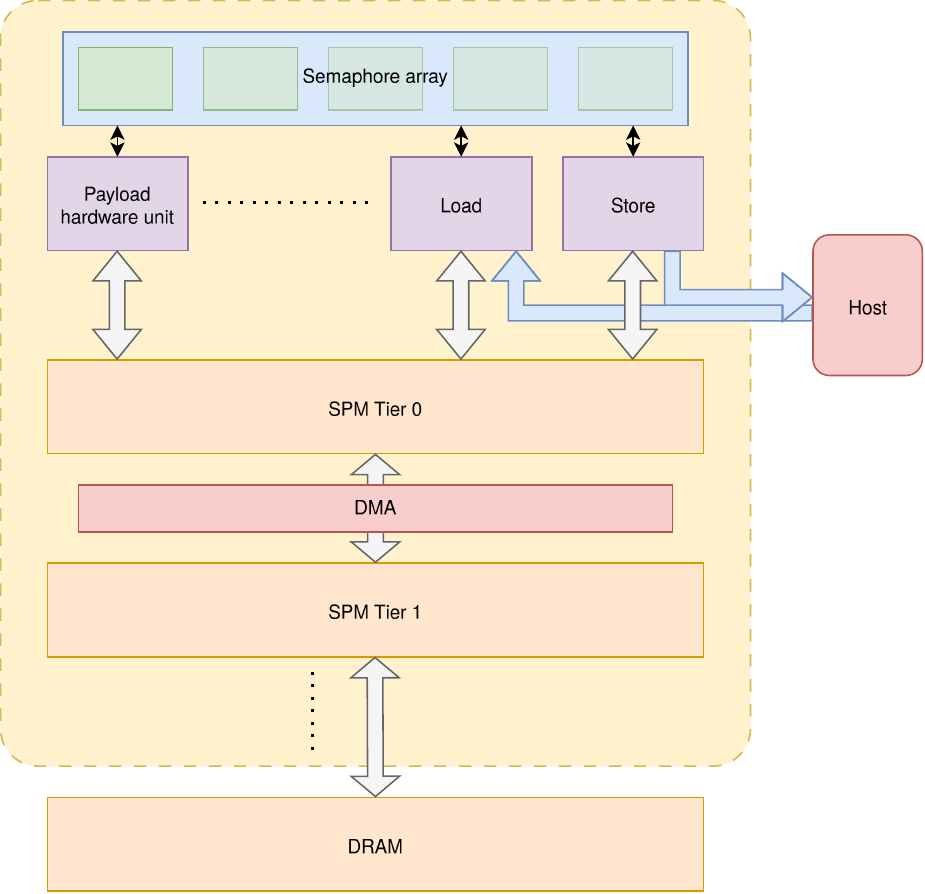}
\caption{Abstract EAAC architecture showing hardware units, tiered memory, and semaphore array for synchronization.}
\label{fig:abstract_arch}
\end{figure}

The system is capable of hosting an arbitrary number of hardware units; all of the hardware units share access to a Tier 0 scratchpad memory for mutual communication. In its default configuration, EAAC provides a Load and Store unit, which provides duplex data transfers to and from a host processor. The input and output data for a given application is written into the memory of EAAC via the Load and Store interfaces. Constants are preloaded to memory via a memory interface exposed to the host.

EAAC organizes its memories into a configurably tiered hierarchy interleaved with programmable DMA units.

Many of the architectural choices made in EAAC promote the use of relatively large, coarse-grained functional units invoked via CISC-like instructions. Examples include various linear-algebra kernels such as GEMM, ReLU, and MaxPool. Because individual instructions represent large units of work, we chose to construct the architecture as a memory-to-memory system---with the Tier 0 SPM being the working memory for the hardware units.

Unlike conventional ISAs, where operands typically represent scalar values or registers, EAAC treats operands as pointers to memory buffers. Hardware units are therefore responsible for interpreting the data contained within these buffers according to the operation being executed. Any operation that can be expressed as a transformation of a finite number of buffers in memory is naturally supported by EAAC; it is up to the user to design a hardware unit capable of the intended operation.


The architecture additionally allows for a degree of out-of-order processing between different hardware units. This design aspect is in line with the general data-driven control of the processor, where a given data-flow node fires when its operands are available, although subject to certain limitations.

EAAC utilizes a system of explicit asynchronous synchronization between the different hardware units to ensure hazard-free data transfers through the shared Tier 0 scratchpad memory. The synchronization primitives chosen are counting or binary semaphore pairs \cite{4b2c511edce04d2ca1ee51d4129c6fa4}. Semaphores consist of two counters in memory representing the empty and full space in a buffer. Counting semaphores, unlike binary semaphores (Mutexes), allow for streaming behavior where full buffers are produced and consumed in smaller portions. Acquiring exclusive access to a buffer or a section of a buffer when streaming, using semaphores necessitates atomic operations when performed asynchronously through a shared memory interface. To accelerate atomic access to the semaphore pairs, EAAC utilizes hardware semaphores.



To acquire access to an input or output buffer needed for some computation, a given hardware unit will attempt to acquire access to the semaphore guarding that specific section of memory. When access is granted by the semaphore, the compute unit can safely access the buffer or a section of it.

Semaphores can exhibit a locked state when both counter values are set to zero; by placing semaphores in this state, we can further control the scheduling of the accelerator units
implicitly. The hardware-semaphore system supports complex chaining behavior, allowing a semaphore to remain in a locked state until another semaphore enters a certain state. This allows for fine-grained fences, as well as broadcasting behavior where a single producer of data can control multiple semaphores, and by extension broadcast data to multiple consumers.


Chaining is managed via the semaphore programming instructions. In the case of broadcasting, semaphore chaining is used to generate multiple consumer semaphores from a single producer semaphore. 


Due to the partial out-of-order execution of the accelerator, as well as the relatively limited number of hardware semaphores available in the system, multiple functional units that do not share a producer/consumer relationship can erroneously attempt to access a specific semaphore pair at the same time. To avoid this issue, the semaphores allow for a degree of renaming that prevents aliasing when multiple hardware units utilize the same physical semaphore pair for different operations. The renaming works by appending a number of ``generation'' bits to the end of the semaphore address passed to the functional units, effectively creating a larger number of virtual semaphores that all map to the same address. A semaphore is programmed with a generation tag; if a functional unit attempts to interact with a semaphore that is not currently holding the same generation tag, access is denied, and the functional unit enters a spin-wait loop. The compiler is responsible for assigning generation bits to both the semaphore and the different operations that access the given semaphores. 


We implement semaphores as dedicated hardware units, rather than, e.g., as atomic operations on an SRAM array, to avoid routing synchronization traffic through main memory and thereby reduce contention and keep synchronization latency predictable.

The semaphore system does not preclude the use of finer-grained, local synchronization mechanisms within individual hardware units. A hardware block may implement internal dependency tracking and use the semaphore array only for communication with DMA units or other hardware blocks.

Explicit asynchronous synchronization adds some overhead in execution time. To minimize this factor, EAAC exploits the fact that its workloads are statically analyzable, pruning the use of semaphores during compilation. Instructions can be issued with or without semaphores guarding access to buffers. Semaphores are initialized using dedicated semaphore-programming instructions.

\subsection{Compiler Design}

The EAAC compiler utilizes a combination of passes from upstream MLIR, as well as passes constructed specifically for EAAC. The passes are performed sequentially and can roughly be partitioned into three categories, in order of execution:

\begin{itemize}
  \item Static memory allocation and data-planning.
  \item Dependency management and hardware semaphore allocation.
  \item Codegen for accelerator units and embedded RISC-V processors.
\end{itemize}

\subsubsection{Memory Allocation}

Bufferization on tensors provides the input IR with explicit allocation and deallocation of buffers used to store intermediate results in memory. EAAC fundamentally works with memrefs representing statically allocated buffers as first-class datatypes. While the EAAC compiler architecture was designed to be workload-agnostic and therefore flexible enough to support different domains, statically allocated memrefs serve as the universal entry point into the compiler.

At this stage in the compilation, the IR consists of operations on memrefs representing statically allocated buffers with well-defined lifetimes. Static memory allocation can thereafter be performed by adding offsets in global memory to the memref buffers, using MiniMalloc \cite{10.1145/3623278.3624752} as the underlying solver. 

EAAC uses MiniMalloc because it is fast, self-contained, and does not depend on an external ILP or CP solver, keeping the compiler's dependency footprint and allocation overhead minimal. The faster execution time allows for fast compilation and, by extension, fast iteration.

EAAC uses a multi-tiered scratchpad memory system. Tier 0 is the only memory visible to the hardware units in the accelerator and functions somewhat like the register file in a conventional CPU. 

\begin{figure}
\centering
\includegraphics[scale=0.55]{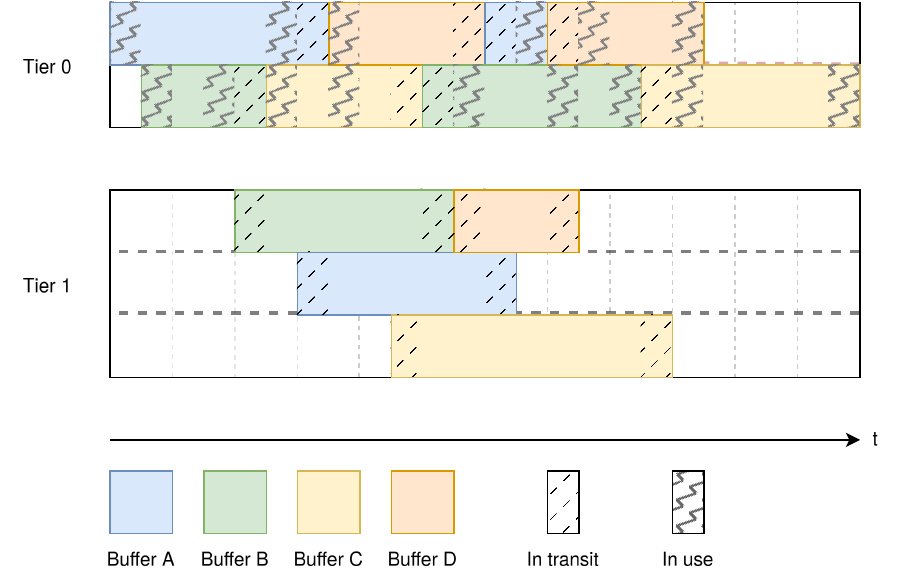}
\caption{Example of buffer spilling. The diagram shows the lifetimes of buffers in memory. In this example, the lifetimes of four buffers are considered: buffers A through D. In this example, the primary memory tier (Tier 0) can fit two buffers at once; the buffers are equally sized. Note that buffers are deallocated after last use}



\label{fig:buffer_planning}
\end{figure}

Data movement scheduling is integrated into the memory allocation algorithm using a modified approach inspired by linear-scan \cite{10.1145/330249.330250}, with spilling used to move data through the hierarchy according to the lifetime requirements extracted from the source program. Whereas the standard use case of MiniMalloc involves creating a problem consisting of all buffers used throughout the execution of the given algorithm, the EAAC implementation sorts the buffers according to the start of their lifetime and grows the problem until MiniMalloc fails due to missing space in a given tier of memory. When it encounters a failure, the algorithm recursively spills the buffer with the latest next use into a lower tier and reloads it into the original tier before its next use. When enough buffers have been spilled to free up space for the next buffer, the algorithm proceeds. The additional spilling logic is implemented as an overlay on top of MiniMalloc within the EAAC compiler; no modifications have been made to the MiniMalloc source code. An example of how buffers are moved between memory tiers is shown in Figure ~\ref{fig:buffer_planning}. 


After static offsets are assigned to the memrefs, a subsequent pass materializes the buffer spills generated by the previous memory-allocation pass into concrete IR operations; these are later lowered into explicit DMA operations to be executed in hardware. When moved between memory tiers, the space previously occupied by a buffer in a given tier is deallocated. Buffers are fundamentally treated as single-assignment and may only exist in a single tier at once except for constants which are statically allocated to the bottom tier of memory.

\subsubsection{Data Dependency Management}

The EAAC compiler is developed to ensure correctness and grow more performant when co-developed with a hardware architecture, as cost-modeling of the hardware improves. In a naive configuration, the compiler has no contextual information about the hardware architecture, and therefore assumes that any given operation executes in a data-driven fashion when its dependencies are available. This implies an arbitrary degree of out-of-order execution, and therefore necessitates a control system that tracks the state of every buffer in scope at any given time. In practice, the compiler manages the state of buffers using virtual semaphores, which are semaphores not assigned to physical hardware semaphores yet. To ensure correctness, the EAAC compiler will allocate a virtual semaphore for every edge between a producer and consumer node in the dataflow graph. This results in a very large number of virtual semaphores, which is infeasible unless an exceedingly large number of hardware semaphores are provided by the hardware architecture.




In the subsequent pass, the compiler analyzes the address space of the accelerator (SPM Tier 0) to identify cases where read-after-write (RAW) errors due to address-space reuse are likely to occur. If the compiler finds an instance of address aliasing within a sliding window, it locates the semaphores guarding the given buffers and chains them in order of use. In the case of RAW, the semaphore guarding the buffer being written is chained to the semaphore guarding the buffer being read, to ensure that the two operations execute in order.


To address the large number of virtual semaphores generated, the compiler will attempt to remove as many of these semaphores as possible, while still ensuring correct execution. 

The specifics of which methods are utilized depends on the architecture of the hardware accelerator targeted by the compiler. When adapted to target the prototype accelerator described in greater detail later, the compiler will eliminate the use of a semaphore between two operations if it can prove that the two operations are implicitly serialized in hardware. Additionally the compiler can eliminate semaphore synchronization if two operations are deemed to have an adequate mutual distance in the data-flow graph, which can be determined from different design parameters in the hardware architecture.



Semaphore optimization can be refined continuously while more detailed hardware modeling is developed.

As the final pass in the data-dependency management section of the compiler, 
virtual semaphores are converted into physical semaphores, by assigning the virtual semaphores to specific hardware semaphores, identified by an address.


The pass scans through the allocated semaphores, sorted in order of allocation time, while keeping a working set of free addresses, and allocates them sequentially. When the algorithm eventually runs out of free semaphore addresses, it recycles a previously used address. To prevent aliasing, the compiler appends a semaphore generation tag to each new allocation, which ensures that access to the underlying physical semaphore occurs in order. As recycling introduces unnecessary serialization, the algorithm always picks the semaphore most likely to be freed soonest, to avoid potential delays. 

The use of generation tags can be compared to register renaming in Tomasulo's algorithm, where multiple versions of the same architectural register can coexist and are distinguished by different tags.

Similarly, generation tags distinguish different allocations of the same underlying semaphore address.


\subsubsection{Codegen}

Code generation is facilitated through a combination of lowering passes expressed in the MLIR Transform dialect \cite{10.1145/3696443.3708922} and more traditional C++ passes. The EAAC compiler takes full responsibility for the orchestration of data movement and synchronization; all compiler passes needed for the management of memories and hardware semaphore synchronization are supplied by EAAC by default. Users are responsible for developing the codegen needed for their specific application. Any operations not natively supported by dedicated hardware units can be offloaded to a RISC-V through functionality provided by the EAAC compiler. In this pipeline, operations are first lowered into the LLVM dialect, which is later compiled into RISV-V assembly. This LLVM lowering is mostly provided by MLIR, but includes some custom compiler passes, namely a compatibility layer that allows the RISC-V processor to access the statically allocated buffers in the scratchpad memory and to interface correctly with the hardware semaphores.

The final output from the compiler is in the form of a FlatBuffer \cite{flatbuffers-github}, which is imported into an assembler and runtime that configures and programs the accelerator. If offloading functionality to a RISC-V processor is necessary, the compiler also generates an .ll LLVM IR file, which is compiled into RISC-V assembly within the compiler pipeline and programmed onto the embedded processor by the same runtime that configures the rest of the accelerator.

\subsection{Hardware Accelerator Prototype}

As a proof-of-concept application of the EAAC design principles, we re-engineered an existing GEMM accelerator \cite{ata8-github} to be compatible with the design elements described above, using it as a prototype target for the EAAC compiler.
The accelerator integrates the core EAAC features described in the preceding sections: a multi-tiered memory system with Tier 0 directly accessible from the main compute units, and a hardware semaphore bank with interfaces to each functional unit. The size of the memories, the number of semaphores, the bus size, as well as the dimensions of the systolic array in the GEMM unit are all parameterizable. 


The functional units are as follows:

\begin{itemize}
  \item A load and store unit that provides fully bidirectional data transfers to and from the host using AXI4-Stream interfaces.
  \item An 8-bit 16x16 GEMM accelerator unit.
  \item A small integer RISC-V processor for various integer workloads such as activation and re-quantization.
\end{itemize}

In the prototype accelerator, instructions are received from a shared decoder. For increased bandwidth, the hardware units feature two memory interfaces: one for interacting with the main scratchpad memories, and another for interacting with the hardware semaphores. In more resource-constrained configurations, these could easily be merged into a single interface by mapping the hardware semaphores into the main address space.

Since data supply and synchronization in EAAC are entirely memory-mapped, the only interfaces needed when integrating a custom hardware unit are a memory interface and a method for receiving instructions from a host.

\begin{figure}
\centering
\includegraphics[scale=0.4]{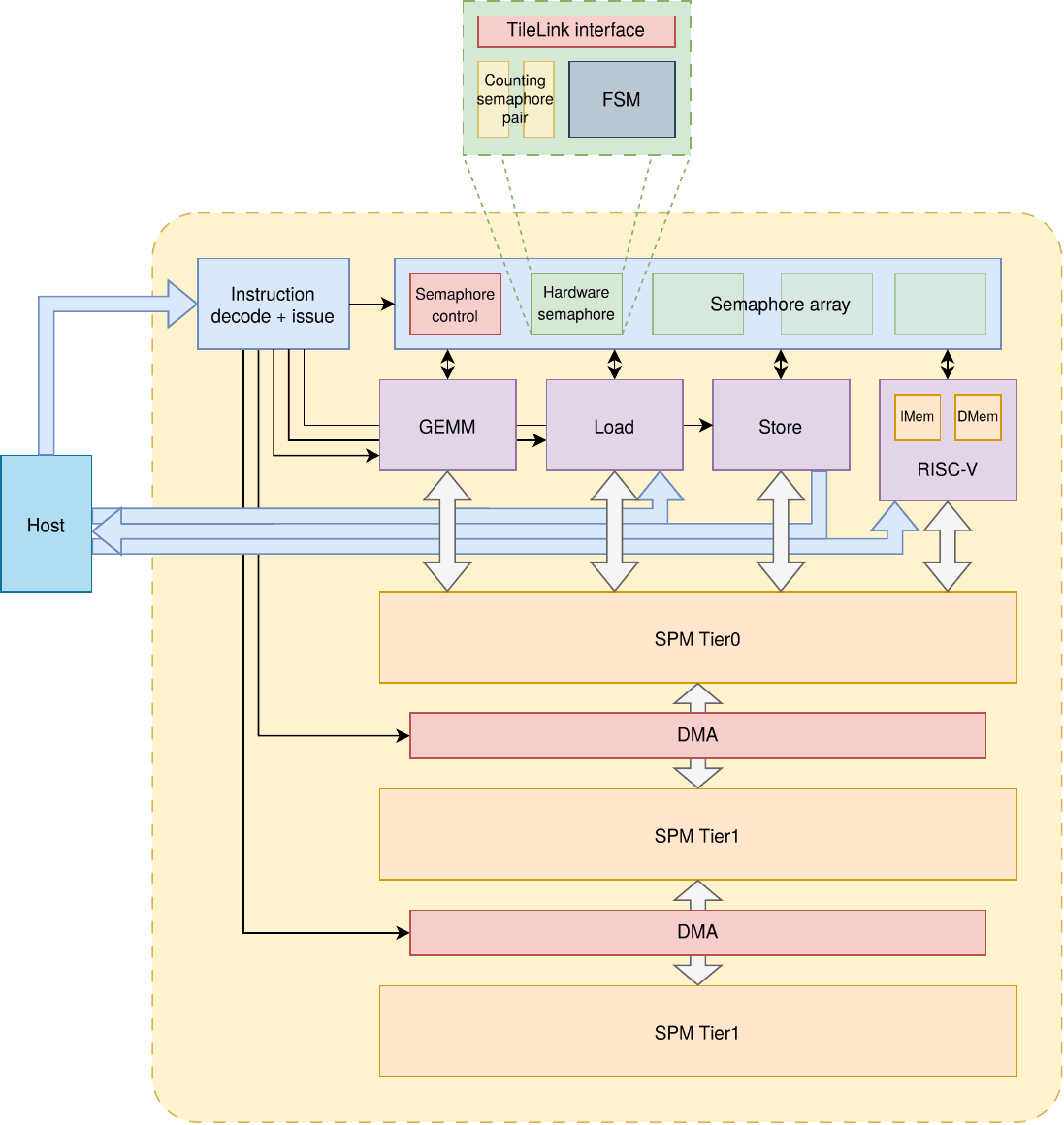}
\caption{Hardware architecture of prototype GEMM accelerator.}
\end{figure}

The system uses a variable-sized bus; the specific implementation uses a 512-bit bus for memory transfers and a separate 16-bit bus for semaphore interaction.

The hardware semaphores are organized into a bank, along with extra logic for the chaining functionality described earlier. When a semaphore transitions from one state to another, an event is generated. These events are logged and can be used to create predicates, such as "semaphore x generated event y". Semaphore initializations are issued when a given set of predicates is satisfied, in a fashion inspired by triggered instructions \cite{10.1145/2508148.2485935}.  


This mechanism also extends to support broadcasting behavior for instruction outputs with a fan-out of more than one. The system utilizes 16 hardware semaphores; each has two parallel memory ports for simultaneous access by both a consumer and a producer of data.

The host interfaces are realized using an AXI4-Stream interface with two full channels allocated for data transfers, and a third used to program the hardware architecture.

The RISC-V processor integrated into the accelerator and later used as the baseline for testing is a custom design based on the five-stage riscv-Sodor pipeline \cite{riscv-sodor-github}. Sodor has been modified with a TileLink data-memory interface that allows it to interact with the hardware semaphores and the main Tier 0 scratchpad memory. Additionally, the processor has been extended to support atomic memory operations through the ``Zaamo'' extension, as well as integer multiplication through the ``M'' extension \cite{riscv-unprivileged-20240411}. The processor is a single-issue, in-order, integer-only processor, and was originally implemented specifically for light control workloads. The ALU computes i32 integer multiplication in a single clock cycle.

The Sodor core on which this design is based is intentionally simplistic, built for educational purposes, and therefore does not offer high performance. It was chosen for its simplicity, to ease integration into the accelerator system; a more performant alternative would be the Rocket Core \cite{Asanovi:EECS-2016-17}, possibly with vector instructions. The RISC-V processor is integrated with small, private data and instruction memories.

\section{Evaluation}

For the evaluation, we ran experimental workloads on the prototype hardware accelerator, both to demonstrate the full-stack integration of the compiler and hardware accelerator and to analyze the execution-time overhead of the data movement and explicit synchronization introduced by EAAC.

Synthesis and area figures for the design were generated using Xilinx Vivado 2025.2, targeting the Kintex 7 xc7k160tffg676-2 FPGA. The baseline for the evaluation is the execution of the workload entirely on the small integrated RISC-V described in the hardware implementation section. 
A second implementation integrates the RISC-V with a 16x16 systolic array GEMM unit.


The target workload is a synthetic dense-layer unit inpired by the MLPerf-Tiny anomaly-detection network. 24 independent $16\times16$ int8 matmul tiles are summed in int32, then bias-added, ReLU'd, and requantized back to int8. 

Larger workloads were pursued but often necessitated a infeasible amount of hardware semaphores due to sub-optimal semaphore optimization in the compiler. For this reason a smaller workload was chosen to not overly inflate the hardware consumption of the hardware semaphore bank.


The system is configured with 16 hardware semaphores, for optimal performance as well as a 64kB of T0 memory, and 128kB of T1 and T2 memory. The mcc core is equipped with 64kB of instruction- memory, and 4kB of data-memory.



\begin{table}
    \centering
    \caption{Execution time breakdown for scaling\_k24: baseline RISC-V implementation vs. hardware accelerator. GEMM and mcc compute cycles overlap in wall-clock time and are not additive; Total reflects the measured wall-clock cycle count.}
    \label{tab:execution_time_scaling_k24}
    \footnotesize
    \begin{tabular}{lr|lr}
        \toprule
        \multicolumn{2}{c|}{\textbf{Baseline RISC-V}} &
        \multicolumn{2}{c}{\textbf{Hardware Accelerator}} \\
        \cmidrule(r){1-2} \cmidrule(l){3-4}
        Segment & Cycles & Segment & Cycles \\
        \midrule
        RISC-V boot+compute & 3,591,489 & EAAC RISC-V boot & 24,711 \\
         &  & GEMM compute & 2,092 \\
         &  & mcc compute  & 102,057 \\
        \midrule
        \textbf{Total} & \textbf{3,591,489} &
        \textbf{Total} & \textbf{127,348} \\
        \bottomrule
    \end{tabular}
\end{table}

Table \ref{tab:execution_time_scaling_k24} shows the total execution time of the workload running on the RISC-V baseline, as well as on the hardware accelerator, which splits the workload between a GEMM unit and the integer RISC-V core.

The execution time assumes all processors are in an uninitialized state, so both figures include the time to transfer the compiled binary into the instruction memory of the RISC-V. This transfer time is significantly longer for the accelerator architecture due to the unoptimized state of the compiler pipeline responsible for lowering the internal EAAC IR to LLVM for execution on the RISC-V.


As shown in Table~\ref{tab:execution_time_scaling_k24}, the vast majority of the accelerator's execution time is taken up by integer operations within the mcc RISC-V core; the GEMM operation itself accounts for only 1.64\% of total execution time, while the accelerator achieves a 28x speedup in total execution time compared to the baseline.

\begin{table}
    \centering
    \caption{Resource utilization by top-level module, reported by xilinx vivado.}
    \label{tab:resource_utilization}
    \renewcommand{\arraystretch}{1.4}
    \resizebox{\columnwidth}{!}{%
    \begin{tabular}{lrrrrr}
        \toprule
        \textbf{Module} & \textbf{LUTs} & \textbf{LUTRAM} & \textbf{FFs} & \textbf{B36} & \textbf{DSP} \\
        \midrule
        GEMM unit & 40{,}384 (62.7\%)  & 1{,}088 (20.0\%)   & 34{,}462 (47.9\%) & 0 (0.0\%)   & 258 (97.0\%) \\
        FrontEnd  & 708 (1.1\%)        & 88 (1.6\%)         & 1{,}068 (1.5\%)   & 0 (0.0\%)   & 0 (0.0\%)   \\
        Load      & 4{,}676 (7.3\%)     & 88 (1.6\%)         & 15{,}484 (21.5\%) & 0 (0.0\%)   & 0 (0.0\%)   \\
        MemSys    & 3{,}118 (4.8\%)     & 0 (0.0\%)          & 2{,}004 (2.8\%)   & 16 (100.0\%)& 0 (0.0\%)   \\
        SemSys    & 4{,}333 (6.7\%)     & 32 (0.6\%)         & 2{,}035 (2.8\%)   & 0 (0.0\%)   & 0 (0.0\%)   \\
        Store     & 4{,}546 (7.1\%)     & 0 (0.0\%)          & 15{,}457 (21.5\%) & 0 (0.0\%)   & 0 (0.0\%)   \\
        mcc       & 6{,}649 (10.3\%)    & 4{,}144 (76.2\%)    & 1{,}381 (1.9\%)   & 0 (0.0\%)   & 8 (3.0\%)   \\
        \midrule
        \textbf{Total} & \textbf{64{,}414} & \textbf{5{,}440} & \textbf{71{,}891} & \textbf{16} & \textbf{266} \\
        \bottomrule
    \end{tabular}}
    \\[2pt]
    \raggedright\scriptsize RAMB18 omitted (0 across all rows). Total is the sum of listed modules; it differs from the tool-reported \texttt{ata8} figure due to hierarchical double-counting in the vendor utilization report.
\end{table}

Table~\ref{tab:resource_utilization} shows the total hardware consumption of the prototype architecture, with the systolic GEMM unit accounting for the large majority of resources used. Being a prototype, the current hardware implementation is not highly optimized. Timing analysis shows that the maximum achievable clock speed of the prototype accelerator is 104.602 MHz, with the critical path lying within the RISC-V core.

\subsubsection{Compiler Evaluation}

The performance of the EAAC architecture is closely tied to the efficiency of the compiler's hardware semaphore allocation. Large workloads with hundreds or thousands of individual instructions place a proportionately larger burden on the semaphore tracking system, since more buffers guarded by semaphores persist for longer throughout the execution of the program. 


To test the efficiency of EAAC's semaphore allocation and pruning, a set of increasingly large programs with worst-case buffer lifetimes was generated.


\begin{figure}
\centering
\includegraphics[scale=0.50]{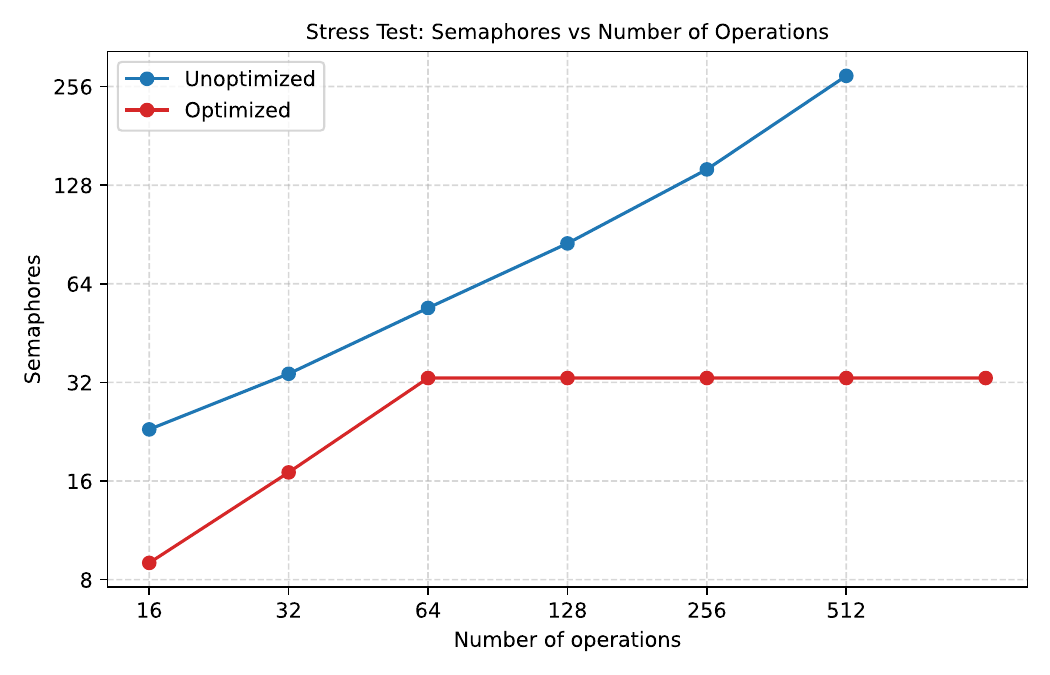}
\caption{Semaphores vs total operations} 
\label{fig:semaphore_scaling}
\end{figure}

As seen in Figure~\ref{fig:semaphore_scaling}, the worst-case number of distinct hardware semaphores needed grows linearly with the number of operations in the application. Since the implemented hardware accelerator utilizes only a single GEMM hardware unit, GEMM operations are naturally serialized, making this number of semaphores entirely unnecessary and therefore low-hanging fruit for optimization.

When leveraging static information about the hardware architecture to optimize semaphore consumption, we are able to break the linear increase in semaphore consumption with total operations. The results show in Figure \ref{fig:semaphore_scaling}, are achieved by eliminating semaphores between operations known to execute in order, as well as eliminating semaphores between operations executing on different hardware units, known to execute in order due to a known degree of run-ahead between hardware units.

%

\section{Conclusion}

We presented EAAC, a compiler and hardware architecture for accelerator co-development that lowers the barrier to integrating custom hardware units into a full software stack. Rather than automating hardware generation, EAAC provides a minimal, extensible interface built around a tiered scratchpad memory hierarchy, static memory allocation via MiniMalloc, and a distributed hardware-semaphore synchronization scheme, allowing accelerators produced through diverse methods to coexist under a shared, memory-mapped control model.

We demonstrated this design through a prototype GEMM accelerator combining a systolic array unit with an embedded RISC-V core. On a synthetic fully-connected-layer workload, the accelerator achieved a 28x speedup in execution time over a RISC-V-only baseline, validating the full-stack implementation from MLIR input through hardware execution.

\section{Acknowledgments}

Claude was used to assist with aspects of code development during the project.
All code was reviewed and verified by the authors, who take full responsibility for the analyses and results.

\bibliographystyle{IEEEtran}
\bibliography{citations}

\end{document}